\documentclass[]{spie}  

\usepackage{amsmath,amsfonts,amssymb, subcaption}
\usepackage{graphicx}
\usepackage[colorlinks=true, allcolors=blue]{hyperref}
\graphicspath{{figures/}}

\newsavebox\IBoxA \newsavebox\IBoxB \newlength\IHeight
\newcommand\TwoFig[6]{
  \sbox\IBoxA{\includegraphics[width=0.45\textwidth]{#1}}
  \sbox\IBoxB{\includegraphics[width=0.45\textwidth]{#4}}%
  \ifdim\ht\IBoxA>\ht\IBoxB
    \setlength\IHeight{\ht\IBoxB}%
  \else\setlength\IHeight{\ht\IBoxA}\fi
  \begin{figure}[!tb]
  \minipage[t]{0.45\textwidth}\centering
  \includegraphics[height=\IHeight]{#1}
  \caption{#2}\label{#3}
  \endminipage\hfill
  \minipage[t]{0.45\textwidth}\centering
  \includegraphics[height=\IHeight]{#4}
  \caption{#5}\label{#6}
  \endminipage 
  \end{figure}%
}

\title{Development and performance of Universal Readout Harnesses for the Simons Observatory}

\author[a]{Jenna E. Moore}
\author[b]{Tanay Bhandarkar}
\author[c]{Brooke DiGia}
\author[d]{Cody Duell}
\author[e,f]{Nicholas Galitzki}
\author[a]{Justin Mathewson}
\author[b]{John Orlowski-Scherer}
\author[d]{Max Silva-Feaver}
\author[g]{Yuhan Wang}
\author[h]{Caleb Wheeler}
\author[i]{Zhilei Xu}
\author[a, j]{Philip Mauskopf}
\affil[a]{School of Earth and Space Exploration, Arizona State University, Tempe, AZ, USA}
\affil[b]{Department of Physics and Astronomy, University of Pennsylvania, Philadelphia, PA, USA}
\affil[c]{Department of Physics, University of California Berkeley, Berkeley, CA, USA}
\affil[d]{Department of Physics, Cornell University, Ithaca, NY, USA}
\affil[e]{Department of Physics, University of California San Diego, La Jolla, CA, USA}
\affil[f]{Department of Physics, University of Texas at Austin, Austin, TX, USA}
\affil[g]{Joseph Henry Laboratories of Physics, Jadwin Hall, Princeton University, Princeton, NJ, USA}
\affil[h]{Center for Computational Astrophysics, Flatiron Institute, New York, NY, USA}
\affil[i]{MIT Kavli Institute, Massachusetts Institute of Technology, Cambridge, MA, USA}
\affil[j]{Department of Physics, Arizona State University, Tempe, AZ, USA}
\authorinfo{Further author information: (Send correspondence to J. E. M.)\\J.E.M.: E-mail: jemoor15@asu.edu}

\begin{document} 
\maketitle

\begin{abstract}
The Simons Observatory (SO) is a ground-based cosmic microwave background (CMB) survey experiment that consists of three 0.5 m small-aperture telescopes and one 6 m large-aperture telescope, sited at an elevation of 5200 m in the Atacama Desert in Chile. SO will utilize more than 60,000 transition edge sensors (TES) to observe CMB temperature and polarization in six frequency bands from 27-280 GHz. Common to both the small and large aperture telescope receivers (LATR) is the 300K-4K Universal Readout Harness (URH), which supports up to 600 DC bias lines and 24 radio frequency (RF) channels consisting of input and output coaxial cables, input attenuators and custom high dynamic range 40K low-noise amplifiers (LNAs) on the output readout coaxial cable. Each RF channel can read out up to 1000 TES detectors. In this paper, we will present the design and characterization of the six URHs constructed for the initial phase of SO deployment.

\end{abstract}


\section{INTRODUCTION}
\label{sec:intro}  
The Simons Observatory (SO) is a new ground-based cosmic microwave background (CMB) survey experiment sited at an elevation of 5200 meters in the Atacama Desert in Chile \cite{sosciencegoals}. SO will utilize three 0.5-meter small aperture telescopes (SATs) and one 6-meter large aperture telescope (LAT) to observe CMB temperature and polarization at both small (degree) and large (arcminute) angular scales. More than 60,000 transition edge sensors (TES) distributed across the four receivers will be used observe the CMB in six frequency bands from 27-280 GHz. These detectors will be read out using microwave SQUID multiplexing ($\mu$-mux) \cite{umux} technology with a targeted multiplexing factor of order 1,000 detectors per radio frequency (RF) input/output pair. Both the total number of detectors and the target multiplexing factor are greater than any previous CMB experiment. \par

\begin{figure}
    \centering
    \includegraphics[width=\textwidth]{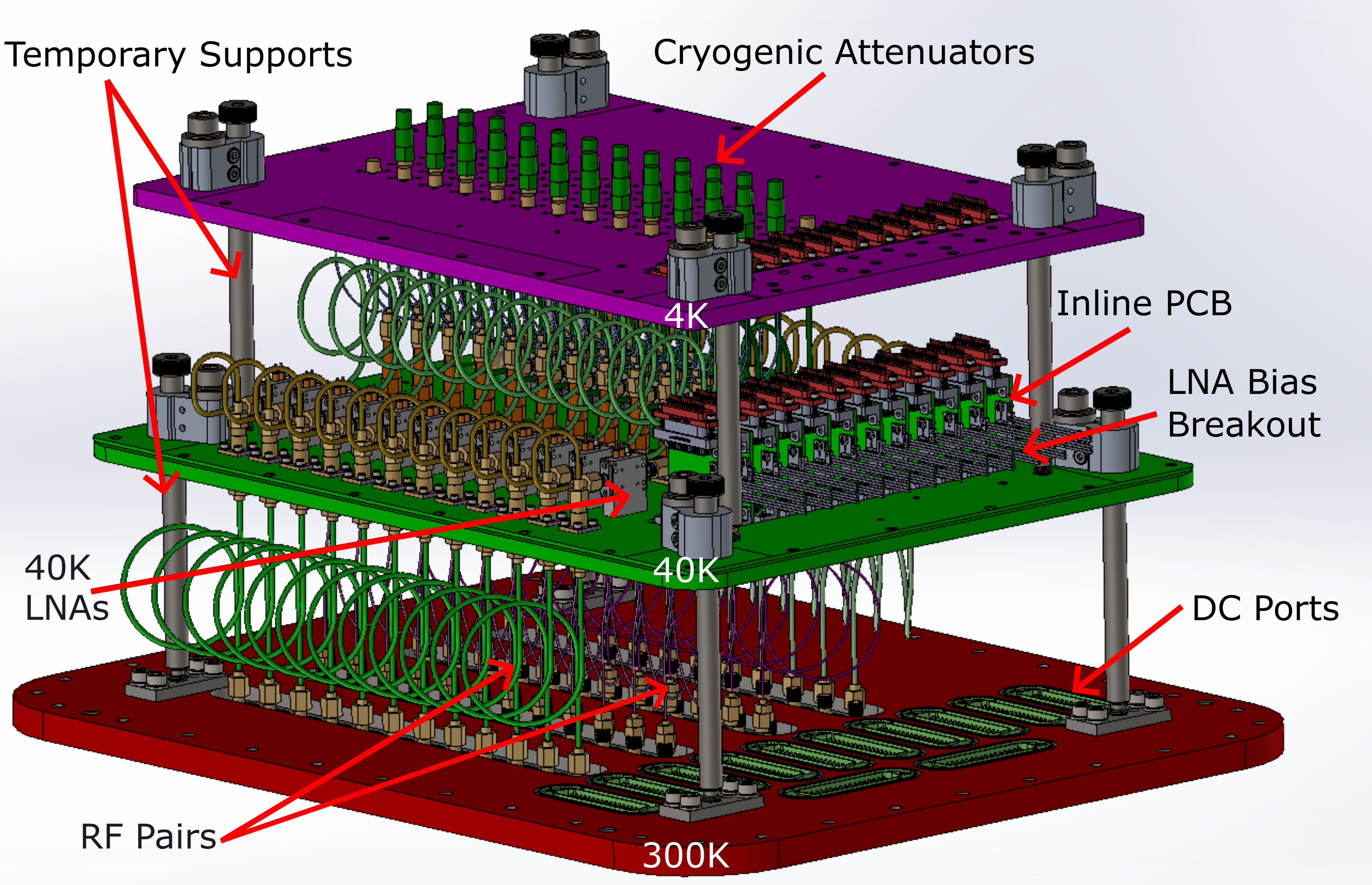}
    \caption{CAD rendering of the Universal Readout Harness. DC ribbon cables are not shown.}
    \label{fig:cadurh}
\end{figure}

\begin{figure}
    \centering
    \includegraphics[width=0.9\textwidth]{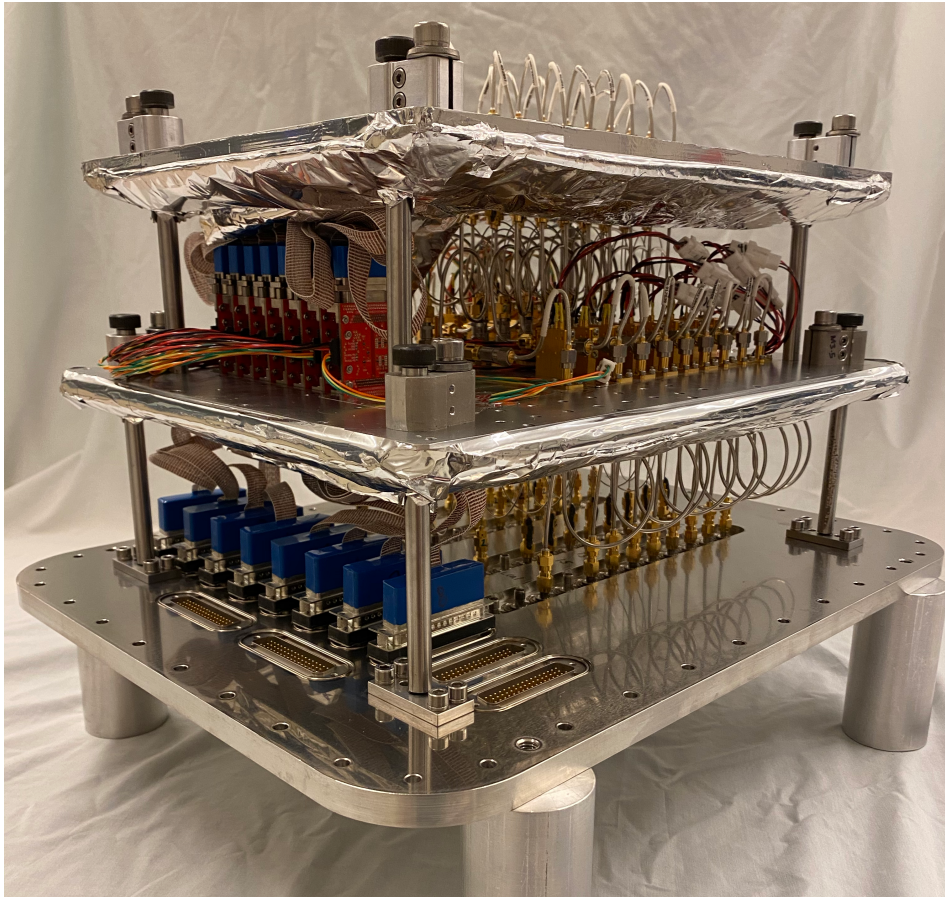}
    \caption{A Universal Readout Harness configured for a SAT featuring 15 RF channels and 400 DC bias lines.}
    \label{fig:irlurh}
\end{figure}

The Universal Readout Harness (URH) (Fig. \ref{fig:cadurh}) is a three-stage assembly that houses all of the 300K-40K and 40K-4K readout wiring and electronics for all Simons Observatory (SO) receivers. A fully populated URH can support up to 24 RF input/output coaxial cable pairs (``RF channels'') and a maximum of 600 DC bias lines grouped together in looms of 50 wires (``DC channels''). With the current SO multiplexing scheme, each RF channel can support the readout of $\sim 1000$ transition-edge sensors (TES). The URH directly interfaces with the SLAC Microresonator Radio Frequency electronics system (SMuRF) \cite{smurf} on the 300K side and with the cold (4K-0.1K) readout on the 4K side. The readout architecture within the URH is standardized across all SO receivers, with the number of populated channels tailored to suit the readout requirements of each receiver. For initial deployment, the small-aperture telescopes (SATs) \cite{sat} each require a single harness with 14 RF channels and 8x50 DC bias lines (Fig. \ref{fig:irlurh}) and the large-aperture telescope receiver (LATR) \cite{latr} requires two harnesses with a collective 46 RF channels. One additional harness is also used in the LATR optics tube testbed (LATRt) \cite{latrt}, which will ultimately be converted into an additional SAT. Housing all of these coaxial cables and components within a standalone wiring harness greatly simplifies the readout assembly and disassembly processes. 

\section{HARNESS DESIGN}

\subsection{Mechanical Design}
\label{sec:mechanicaldesign}
Structurally, the URH is comprised of three aluminum plates held in place by support rods. The aluminum support rods are required only for assembly and installation, and are removed once the harness plates have been securely bolted to the 300K, 40K, and 4K receiver shells. The 300K vacuum plate features hermetically sealed laser-welded SMA\footnote{Ceramtec, Laurens, SC 29360 USA} and 50-position subminiature-D (D-sub)\footnote{Accu\-Glass Products, Inc., Valencia, CA 91355 USA} connectors, while standard bulkhead connectors are used on the 40K and 4K stages. Each URH stands approximately 35x50x270 cm, weighs approximately 25 kg and can be fitted with either forklift tubes or handles to simplify handling and installing into the receiver. \par

Careful consideration must be taken to minimize thermal loading from the large number of components packed into the small footprint of the URH. Low thermal conductivity material is used in cables that connect temperature stages, and attenuators are used to help manage heat loads on the input RF lines. Fixed components such as low-noise amplifiers (LNAs) and PCBs are attached to the harness plates with conical spring washers to ensure good thermal contact is maintained.  Multi-layer mylar insulation (MLI)\footnote{RUAG Space GmbH, 1120 Vienna, Austria} blankets are used on the warm-facing sides of the 40K and 4K harness plates and around the DC ribbon cables. Receiver heatstraps are connected directly to the 40K and 4K harness plates. \par

\begin{figure}[h]
    \centering
    \includegraphics[width=0.85\textwidth]{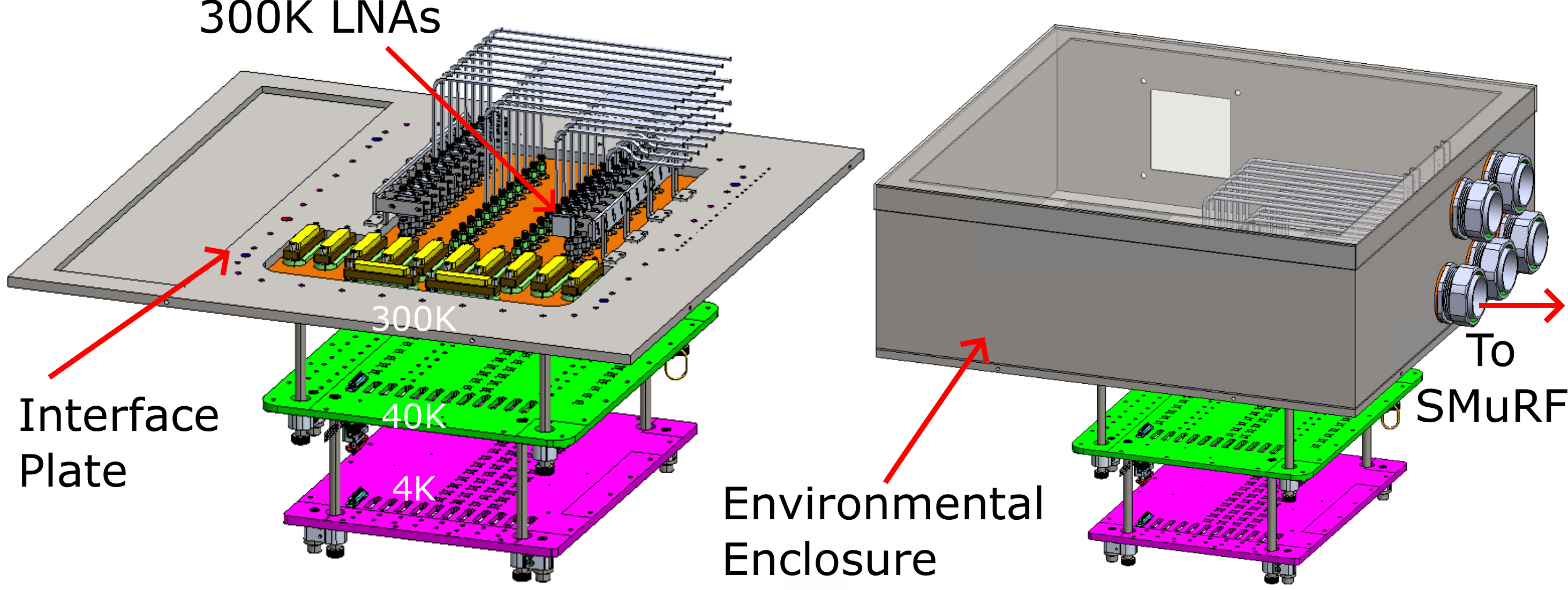}
    \caption{After the URH is integrated into the receiver, warm readout electronics are mounted directly to the air-facing side of the 300K URH plate. An environmental enclosure protects the URH and warm electronics from the elements. Design shown is for the SATs.}
    \label{fig:satenclosure}
\end{figure}

While the URH design and implementation is identical for all SO receivers, the interfaces and integration procedures differ between the SATs and the LATR. In the more densely populated LATR, the 4K stage of the URHs are connected to the optics tubes (OTs) via isothermal coaxial cable ``highways'' that run along the inside of the receiver shell to the OTs. In the SATs, the URH interfaces with the cold readout assembly (CRA) via isothermal coaxial cables distributed radially around the CRA. The endpoint of the cold readout for all SO receivers is the Universal Focal Plane Module (UFM) \cite{ufm}, which contains the multiplexing module, detector stack, and optical coupling. On the warm side, the URH interfaces with SMuRF through a series of warm electronics that are connected directly to the air side of the 300K URH vacuum plate. An environmental enclosure mounted to the cryostat shell protects the air-facing side of the URH and warm readout electronics from the elements (Fig. \ref{fig:satenclosure}). More details regarding the URH interfaces in the different receivers may be found in Refs.~\citenum{latr, satpaperinprep}. 

\subsection{RF and DC Design}
\label{sec:rfdcdesign}
\begin{figure}[b]
    \centering
    \includegraphics[width=0.95\textwidth]{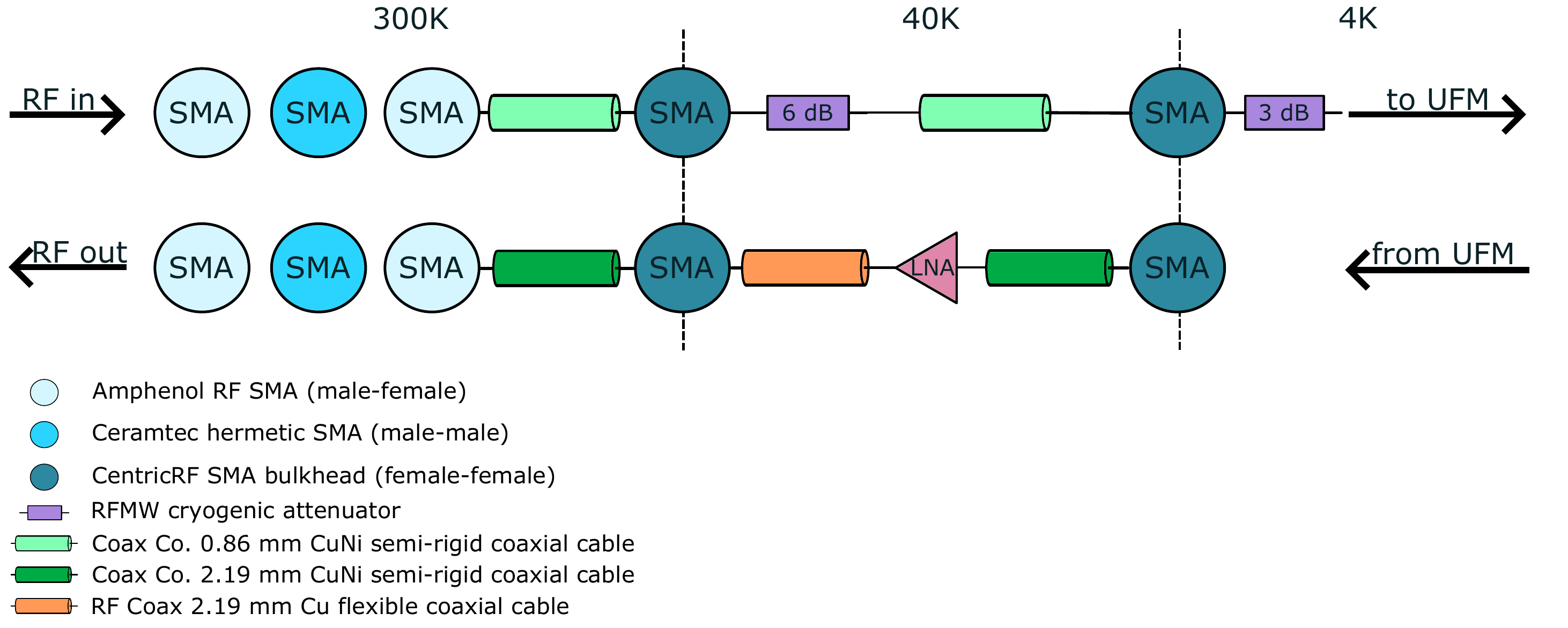}
    \caption{URH RF channel layout. During harness testing, additional pieces of copper coax (not shown) are installed between the 4K attenuators and the 4K SMA bulkheads to bridge RF input to RF output. }
    \label{fig:rfchain}
\end{figure}

A block diagram of the full URH RF chain is illustrated in Figure \ref{fig:rfchain}. Due to its relatively low attenuation and thermal conductivity, semirigid coaxial cables\footnote{COAX JAPAN CO., LTD., Kanagawa 225-0012 Japan} fabricated out of cupronickel (CuNi) are used for connections between temperature stages. Selected cable geometries feature lengths and diameters that strike a balance between high signal-to-noise (SNR) and low thermal conductivity while helical shapes enable dense packing and provide strain relief. For isothermal connections, flexible copper coaxial cables\footnote{RF COAX INC., Hollister, CA 95023 USA} are used. To help moderate input signal levels and reduce thermal noise, 6 dB and 3 dB cryogenic attenuators\footnote{RFMW Ltd., San Jose, CA 95119 USA} are used on the RF input lines at 40K and 4K, respectively. All components are selected to have a characteristic impedance of 50 ohms to minimize reflections due to impedance mismatch. \par

\begin{figure}[!htp]
    \centering
    \includegraphics[scale=0.075]{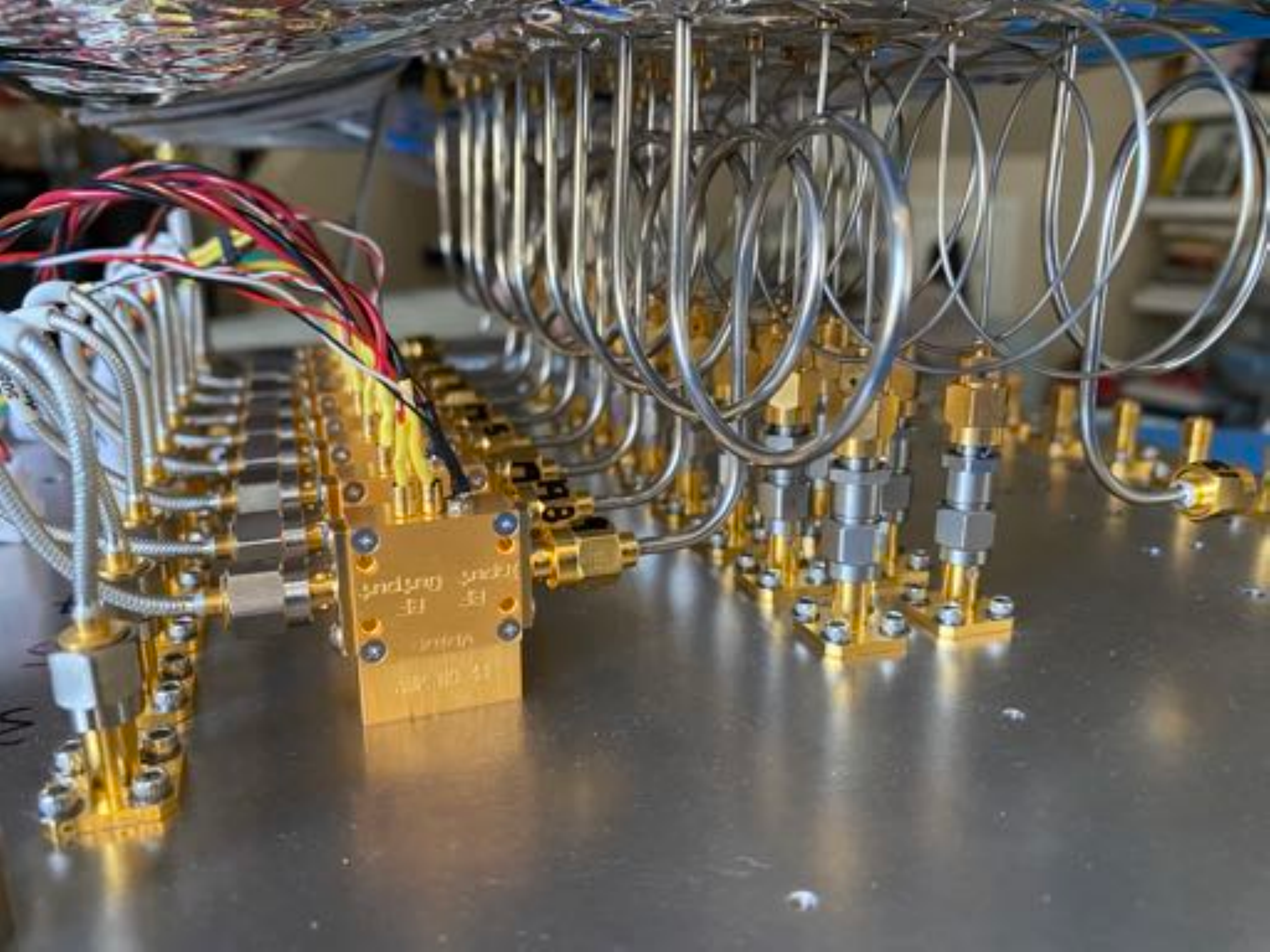}
    \caption{Low-noise amplifiers on the 40K stage of the Universal Readout Harness}
    \label{fig:lna}
\end{figure}

\begin{figure}[!hbp]
    \centering
    \includegraphics[scale=0.45]{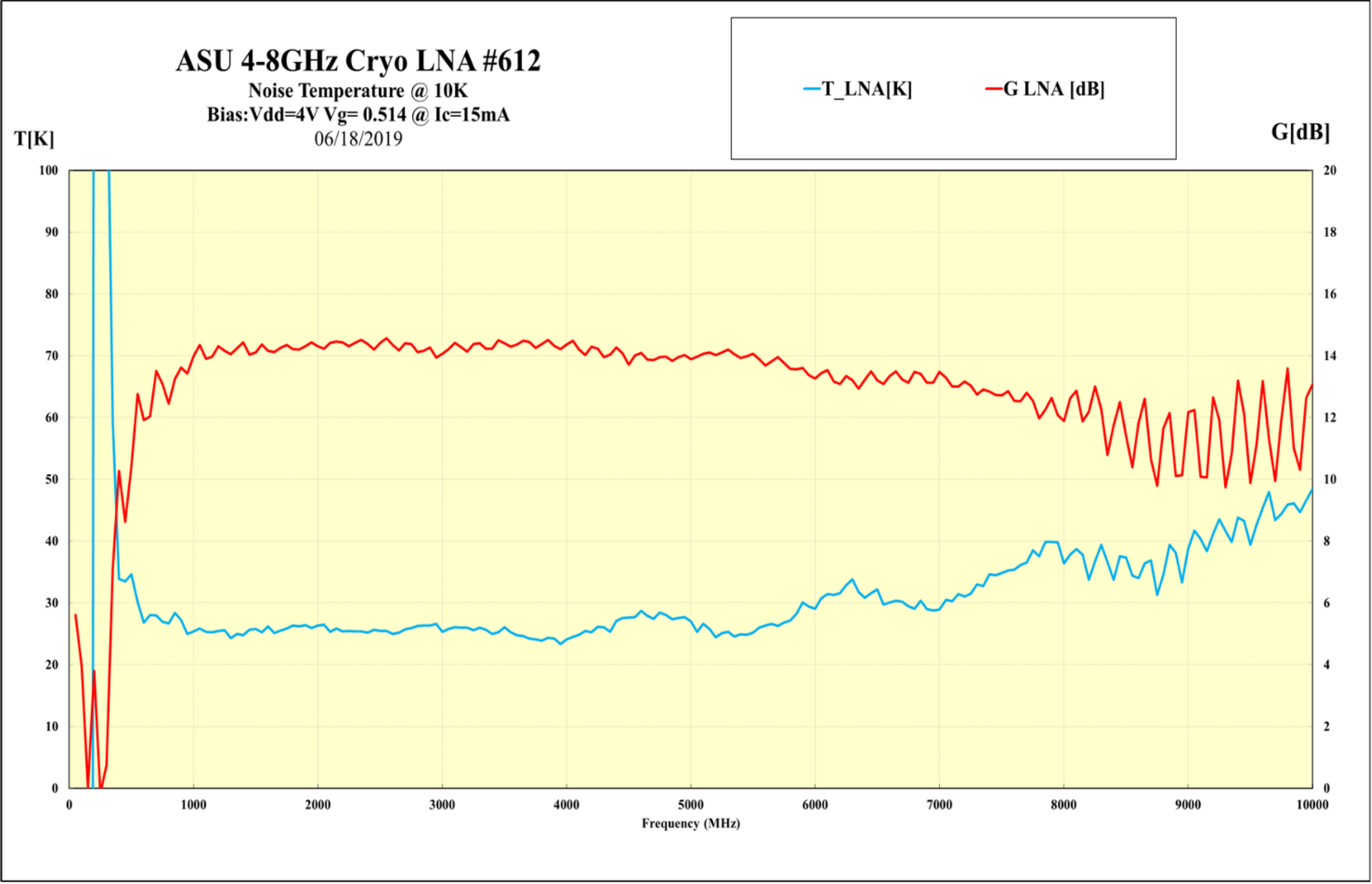}
    \caption{Noise temperature and gain of the 40K low-noise amplifiers measured at 10K.}
    \label{fig:noisetemp}
\end{figure}

A primary feature of the Universal Readout Harness are the custom high dynamic range low-noise amplifiers (LNAs)\footnote{Groppi Labs, Arizona State University, Tempe, AZ 85287 USA, \url{www.thz.asu.edu}} used on each RF output line on the 40K stage (Fig. \ref{fig:lna}). Designed and manufactured in-house specifically to meet the readout noise and linearity requirements for SO \cite{sorfdesign}, they provide flat 14 dB gain across the 4-8 GHz band. A full description of LNA parameters is given in Table \ref{tab:lnaparams} and a plot of amplifier gain and noise temperature is shown in Figure \ref{fig:noisetemp}. The amplifiers feature a three-wire (gate voltage $V_G$, drain voltage $V_D$, and ground) biasing design and receive power from the 40K inline PCBs via breakout wire triplets. \par

\begin{table}[!hp]
    \centering
    \begin{tabular}{|c|c|c|c|c|c|} \hline
        \textbf{LNA parameter} & \textbf{Gain (dB) }& \textbf{IIP3 (dBm)} & \textbf{IP1dB (dBm)} & \textbf{$T_{noise}$ (K)} & \textbf{$P_{diss}$ (mW)} \\ \hline
        \textbf{Value at 4V/15mA/40K} & 14 & 7 & -2 & $\sim30$ & 60 \\ \hline
    \end{tabular}
    \caption{40K LNA parameters}
    \label{tab:lnaparams}
\end{table}

Each URH can support up to 600 DC wires for detector biasing, amplifier biasing, and flux ramping. These wires are grouped together in custom ribbon cable looms\footnote{Tekdata Interconnections Limited, ST1 5SQ Staffordshire UK}. Each cable loom consists of 36 AWG manganin (PhBr) wires in twisted pairs and triplets woven together with DuPont Nomex fiber. On the 4K and 40K stages these cables are terminated in 51-position micro-D (MDM) connectors\footnote{Omnetics Connector Corporation, Minneapolis, MN, 55432 USA} while 50-position subminiature-D (D-sub) connectors are used for the 300K terminations. All connectors are encapsulated with stycast to provide strain relief. On the 40K stage, custom inline PCBs provide power to the 40K LNAs via a 6-pin breakout connector, with each PCB powering two LNAs. This breakout power is carried to the LNAs via 24 AWG twisted triplet pigtails with Molex Ditto genderless connectors. \par

One particular challenge with the URH RF design is the high density of SMA connections. Each RF input/output pair requires a total of 17 SMA connections, each of which could introduce problems if not properly connected. Improper SMA mating due to operator error or faulty connector geometry (Fig. \ref{fig:badcoax}) can not only cause physical damage to the components themselves, but may also result in undesirable features in the microwave signal such as discontinuities in transmission (Fig. \ref{fig:smadip}) or excessive reflections due to  poor contact between the two connector interfaces. If the mismatch is slight, these effects may not be seen until after one or more thermal cycles. Since the densely packed nature of the instrument prevents simple debugging of connections post-assembly, it is imperative that SMA connections are verified at multiple points in the harness assembly process. This is done by verifying the geometry of each SMA connector prior to installation using an SMA gauge kit and using a vector network analyzer (VNA) to examine the s-parameters of components after they are installed in the harness whenever possible. If the s-parameters appear nominal in static and slightly stressed (e.g. applying gentle pressure to semirigid coax loops, wiggling SMA connectors) conditions, the connections are then fixed in place with a thin stripe of ScotchWeld epoxy according to the procedure outlined in Ref.~\citenum{almamemo}. To maintain the integrity of the welded SMA connectors on the 300K plate, epoxy is not applied directly to those connectors. Instead, the semirigid cables are connected to male to female SMA ``connector savers.''

\TwoFig{coax_protrusion.pdf} 
     {A faulty SMA connector. Excess dielectric material protrudes beyond the metal contact ring and prevents proper mating.} 
     {fig:badcoax} 
     {s21_dip.pdf} 
     {An example plot of RF transmission (S21) through a URH RF channel with a faulty SMA connection. The dip in S21 near 5.5 GHz is a result of the poor connection.} 
     {fig:smadip} 

\section{TESTING}
\label{sec:testing}

Each URH undergoes extensive testing in a dedicated test cryostat (Fig. \ref{fig:maneater}) at ASU prior to integration in SO receivers. The primary goal of this pre-integration testing is to verify that the RF chains are performing nominally and to provide baseline 40K LNA bias voltage requirements and performance data. Establishing these baseline performance measures for the URH portion of the SO readout scheme enables tracking of channel/amplifier performance over time and allows for easier debugging in the event that issues arise after integration. This verification is done by performing a series of ``loopback'' measurements, where each RF input chain is connected to its corresponding RF output chain at the 4K stage by a small piece of flexible copper coax. An external power supply connected to the DC ports on the air-facing side of the 300K plate provides power to the 40K LNAs. 

\TwoFig{harness_cryostat.pdf} 
     {Universal Readout Harness installed in the dedicated harness test cryostat} 
     {fig:maneater} 
     {radshield.pdf} 
     {Radiation shields cover the 40K and 4K stages of the harness test cryostat} 
     {fig:radshields} 

\subsection{RF Validation}
Prior to cryogenic testing of each harness, loopback s-parameter measurements are taken at room temperature for each complete RF channel. This serves as a final check of all connections and establishes the baseline shape of the signal through each chain. The harness is then installed in the test cryostat and cooled with a two-stage Gifford-McMahon (GM) cryocooler to approximately 6K (4K stage) and 60K (40K stage). It is suspected that excessive radiative thermal loading from the large radiation shields (Fig. \ref{fig:radshields}) covering the 40K and 4K stages of the test cryostat are the limiting factor in cooling power in this test setup, as thermal analyses conducted with the URH installed in the receivers have indicated the total heat load from the URH to be approximately 7 W at 40K and 0.15 W at 4K, which is within the design requirements \cite{latr}. \par

\begin{figure}[h]
\centering
\begin{subfigure}{.5\textwidth}
  \centering
  \includegraphics[width=.9\linewidth]{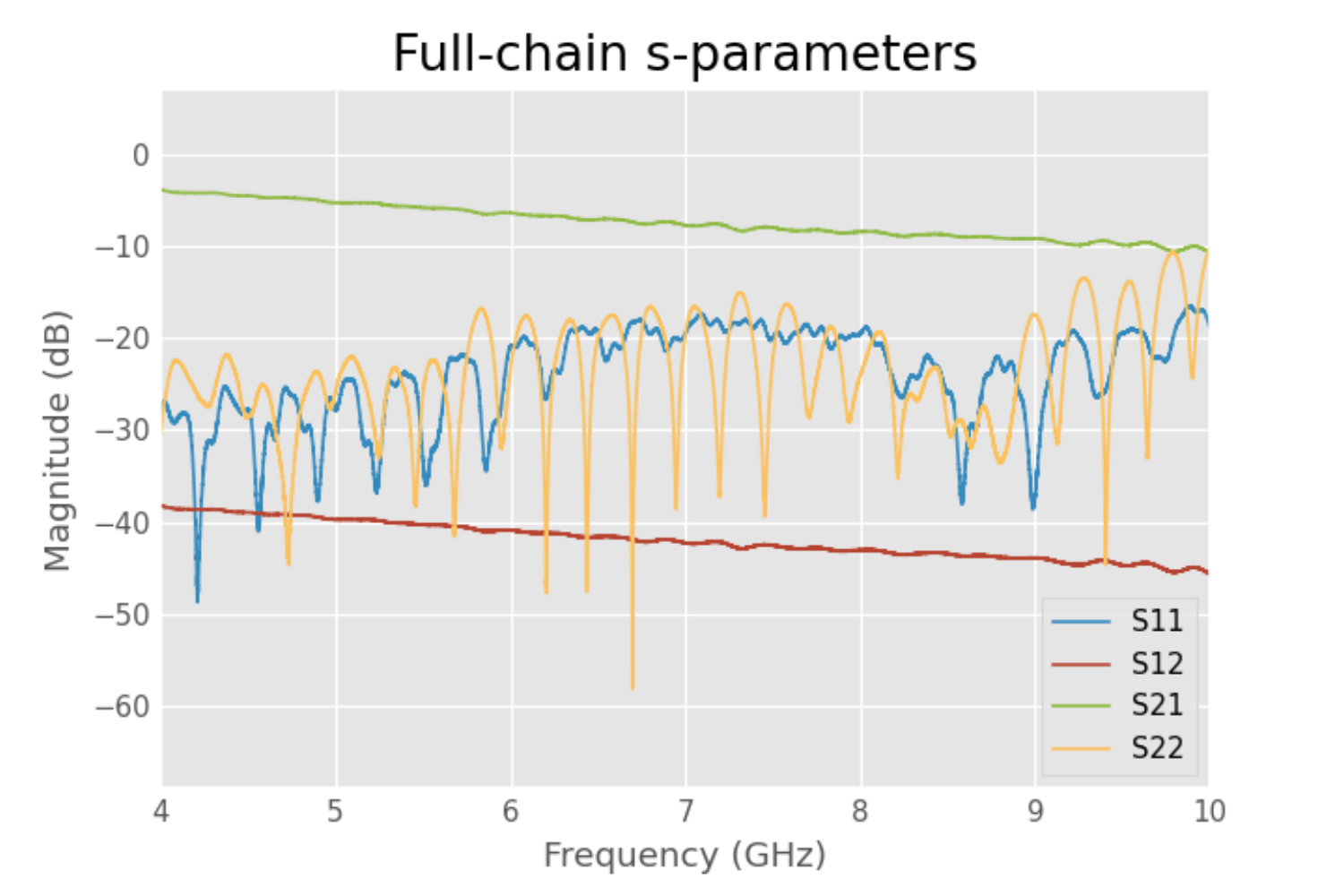}
  \label{fig:fullsparams}
\end{subfigure}%
\begin{subfigure}{.5\textwidth}
  \centering
  \includegraphics[width=.9\linewidth]{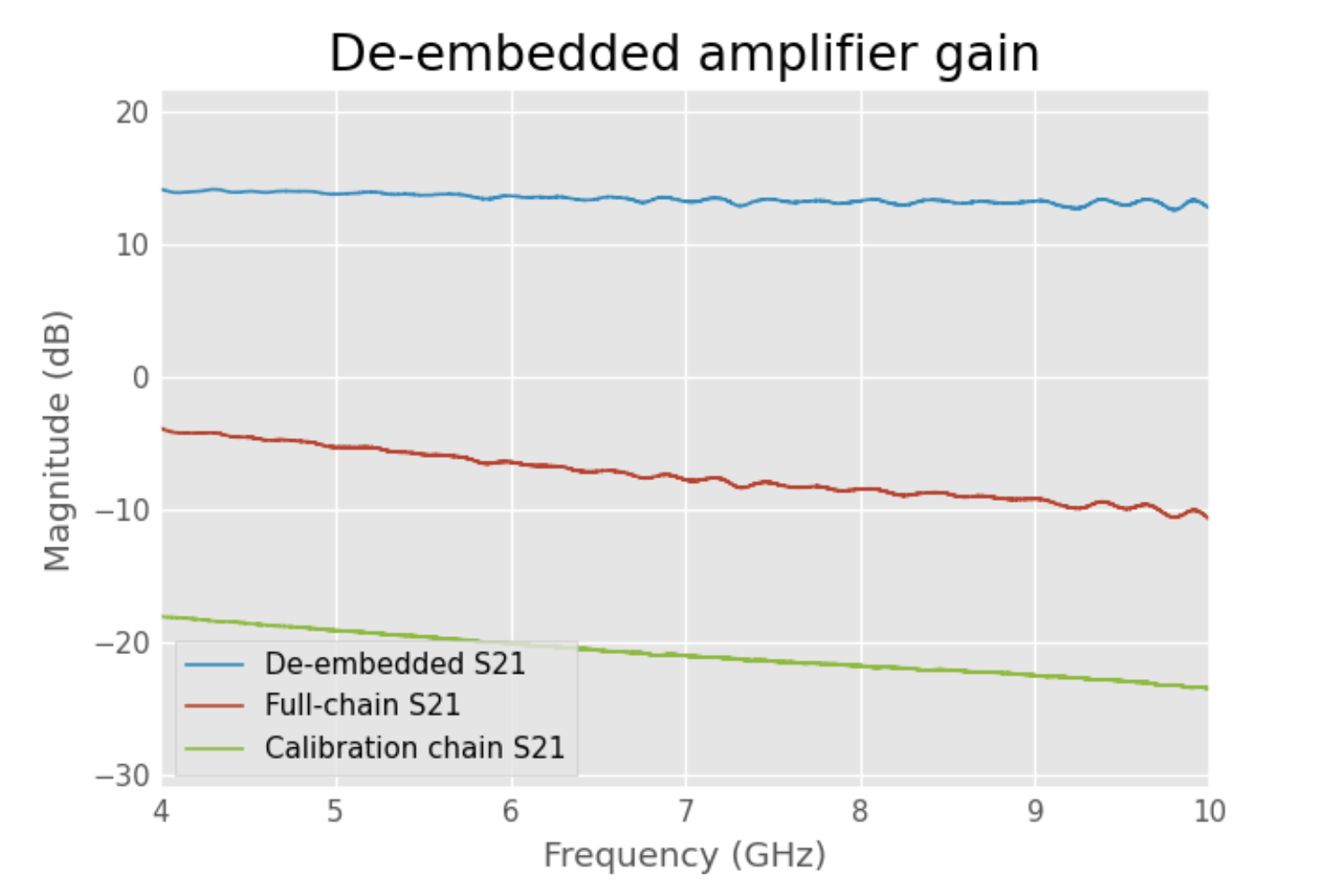}
  \label{fig:deembedded}
\end{subfigure}
\caption{Cryogenic s-parameter measurements of a typical URH RF channel in loopback mode.}
\label{fig:sparams}
\end{figure}

Once the harness reaches base cold temperatures, s-parameter measurements are repeated for each RF channel. These measurements are compared against the 300K measurements to detect any changes in transmission profile shape (e.g. introduction of ``dip'' features). These measurements are saved and provided together with individual amplifier bias parameters to receiver institutions as a reference for how each RF channel should perform in cryogenic conditions. To recover de-embedded amplifier gain measurements, s-parameter measurements of calibration RF chains with a 0 dB SMA thru in place of the LNA are taken. The de-embedded amplifier gain is taken as the difference between the full chain and calibration chain transmission (S21). Due to the scale of the project ($N_{LNA}\sim 100$), de-embedding is preferable to individually cooling and testing each amplifier in a separate testbed prior to installation in the URH. An example of full-chain and de-embedded measurements is shown in Figure \ref{fig:sparams}. \par

Following cryogenic testing, one last set of s-parameter measurements is taken at room temperature after the harness has been removed from the test cryostat. This last set of measurements is used to detect any changes that may have occurred during the thermal cycle and serves as a pre-shipping baseline. These measurements are repeated immediately after unpacking the harness at the receiver institution to ascertain if any damage occurred in transit.

\subsection{Linearity Measurements}
\begin{figure}[t]
\centering
\begin{subfigure}{.5\textwidth}
  \centering
  \includegraphics[width=.9\linewidth]{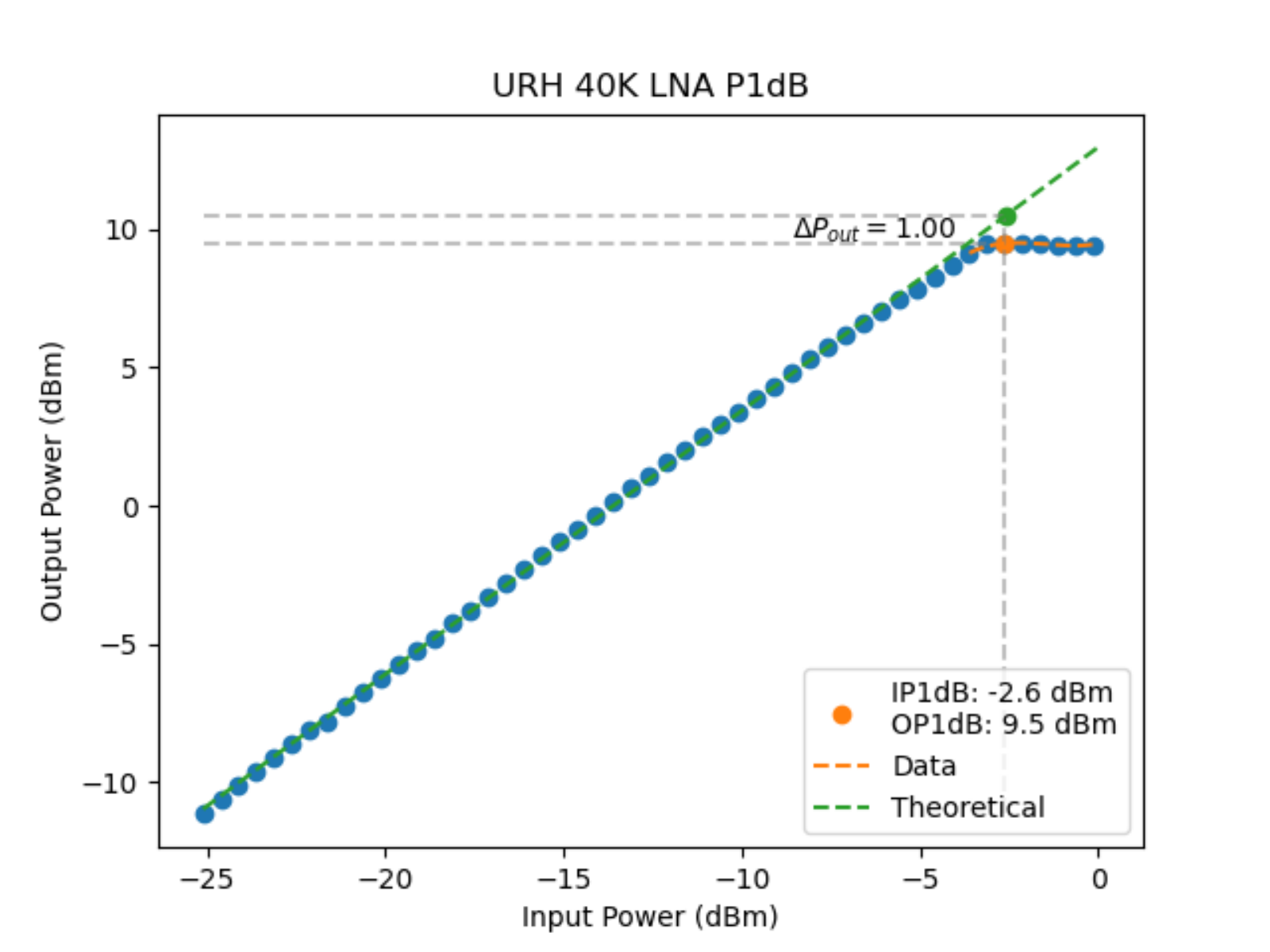}
  \label{fig:p1db}
\end{subfigure}%
\begin{subfigure}{.5\textwidth}
  \centering
  \includegraphics[width=.9\linewidth]{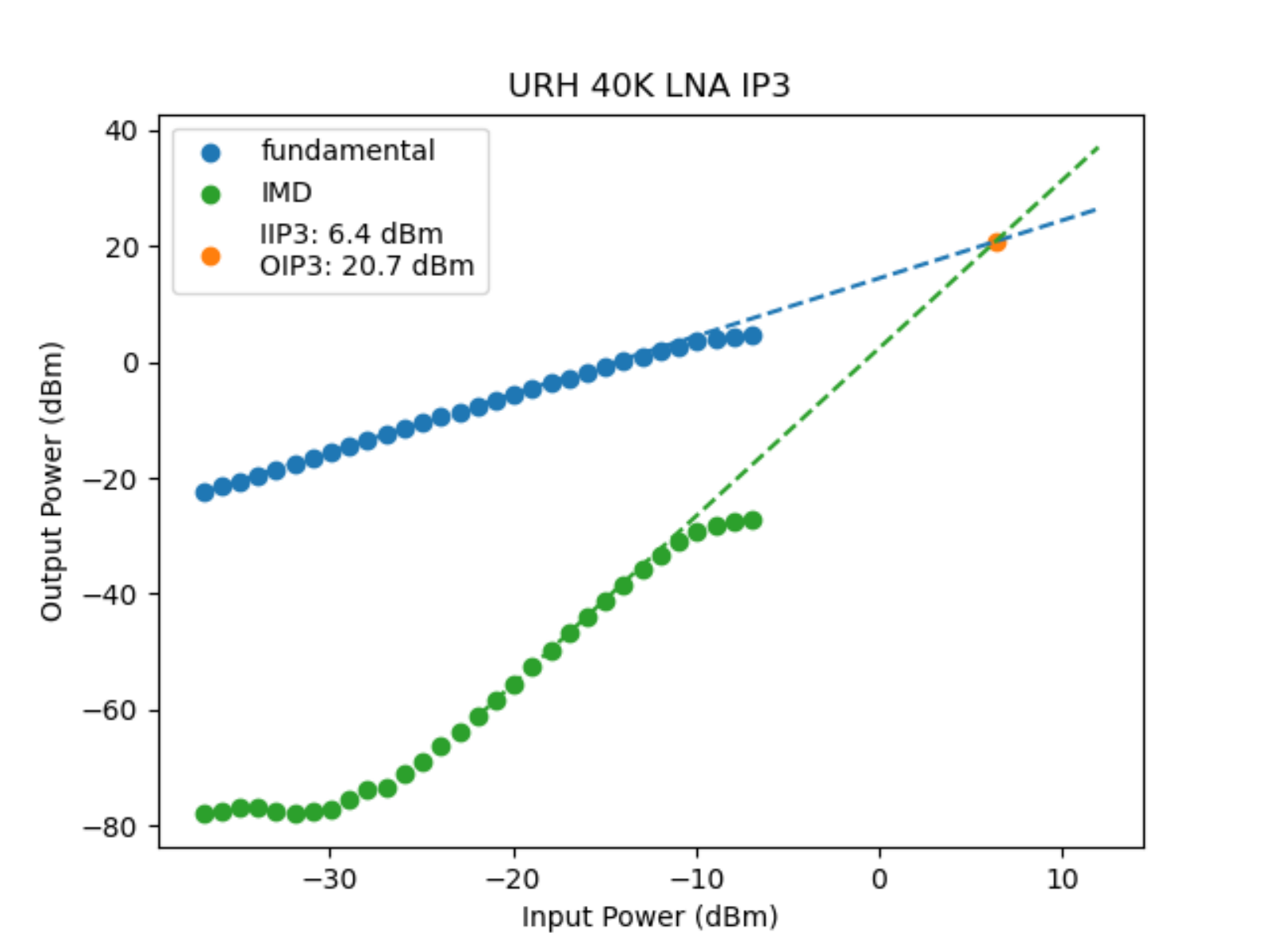}
  \label{fig:ip3}
\end{subfigure}
\caption{In-situ cryogenic linearity measurements of a typical URH 40K LNA. Output power level is plotted as a function of input power level in both plots. Attenuation from additional RF components in between the measurement plane and the LNA input/output has been subtracted out. The plot on the left shows the 1-dB compression point of the 40K LNA. The plot on the right shows the response of the 40K LNA at fundamental and intermodulation product frequencies and extrapolates out to the third order intercept point (IP3)}
\label{fig:linearity}
\end{figure}

In addition to s-parameter measurements, cryogenic measurements of amplifier 1-dB compression point (P1dB) and third-order intercept point (IP3) are also conducted to ensure that each 40K LNA meets the linearity requirements of the SO readout scheme. These tests are also done in-situ, with the attenuation of all non-LNA RF components carefully accounted for and subtracted out in post. While some VNAs have features that allow for simple P1dB and IP3 calculations, the high linear range of the LNAs coupled with the large amount of attenuation in the RF chains ($\sim$18 dB prior to LNA input) means that the input power must be swept up to $\sim$ 20 dB to reach the nonlinear regime of the LNAs, which is beyond the capabilities of available VNAs. Instead, P1dB measurements are done by conducting a single 6GHz tone power sweep with a signal generator connected to the 300K air-facing RF input port. The output power is measured by a power meter connected to the 300K air-facing RF output port, and a Python script is used to fit the response and determine the P1dB. For IP3 measurements, a power combiner is used to combine signals from two signal generators and feed them in to the 300K RF input port. A power sweep of 6 GHz and 6.01 GHz tones is conducted and the response is measured by a spectrum analyzer connected to the 300K RF output port. A Python script fits the response at the fundamental and intermodulation product frequencies and extrapolates to find the intercept point. Example linearity measurements are shown in Figure \ref{fig:linearity}. While these in-situ methods are not as precise as individually cooling and testing each LNA in an isolated environment, the values obtained are good enough to determine whether a given LNA meets the linearity requirement. \par

\section{Conclusions and Future Work}
We have presented the design, construction, and testing of 300K-4K Universal Readout Harnesses for the Simons Observatory. To date, six Universal Readout Harnesses have been assembled, tested, and delivered to collaborating institutions in preparation for the initial phase of SO operation. In addition the six required for operation, a seventh harness is also in development. This ``spare'' harness will be used to further characterize the electrical and thermal performance of the URH and serve as an emergency back up in the event of a major failure. Additionally, this harness will be used to test the in-situ performance of a new proprietary 4K low-noise amplifier recently developed at ASU. While current SO readout systems utilize commercially available 4K LNAs, future iterations and expansions to SO hope to utilize this new design. \par

\begin{figure}[h]
    \centering
    \includegraphics[width=\textwidth]{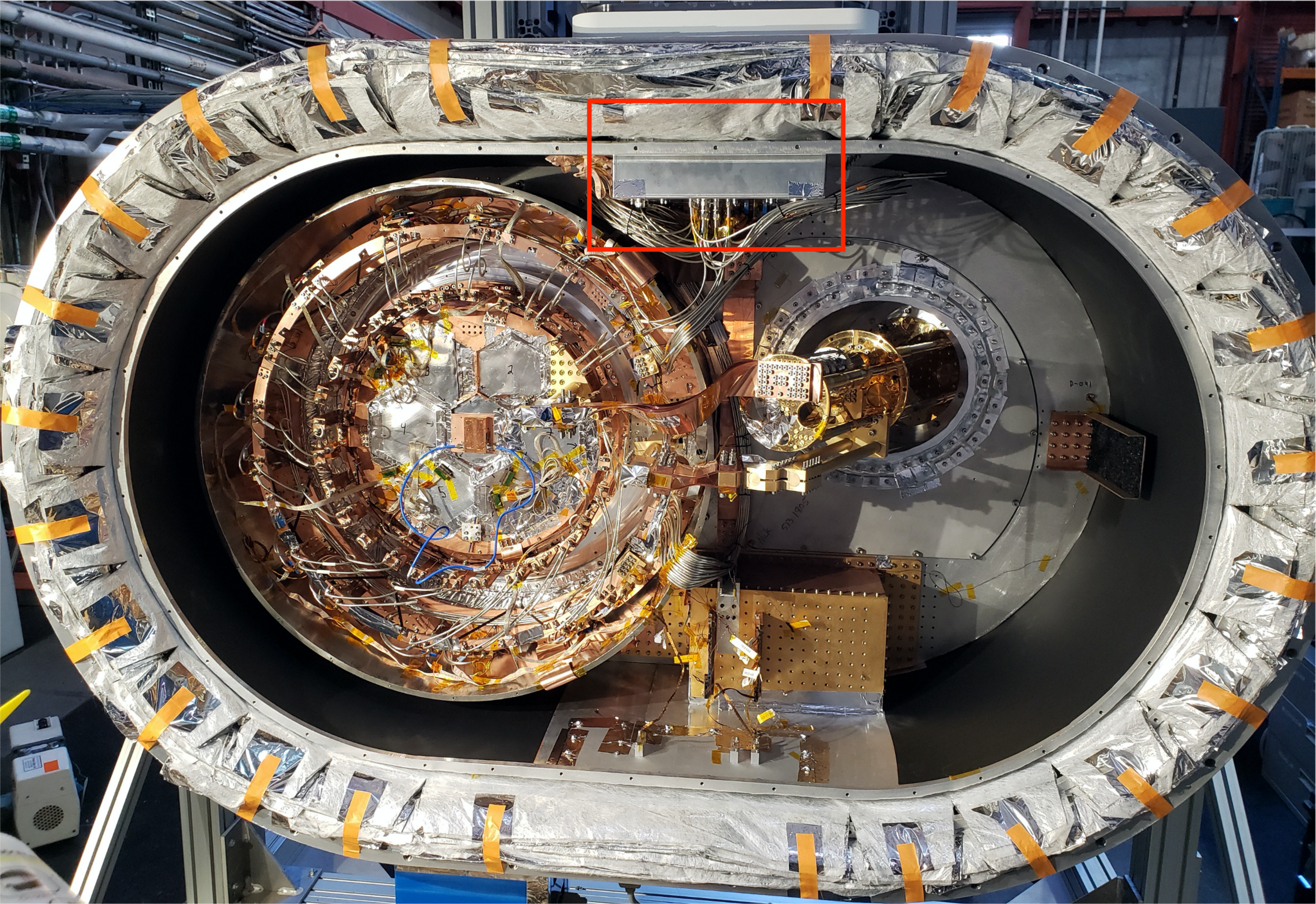}
    \caption{The SAT-MF1 receiver. The integrated URH is outlined in red. Photo courtesy of the Arnold Lab at UCSD.}
    \label{fig:sat1int}
\end{figure}

\begin{figure}
    \centering
    \includegraphics[width=\textwidth]{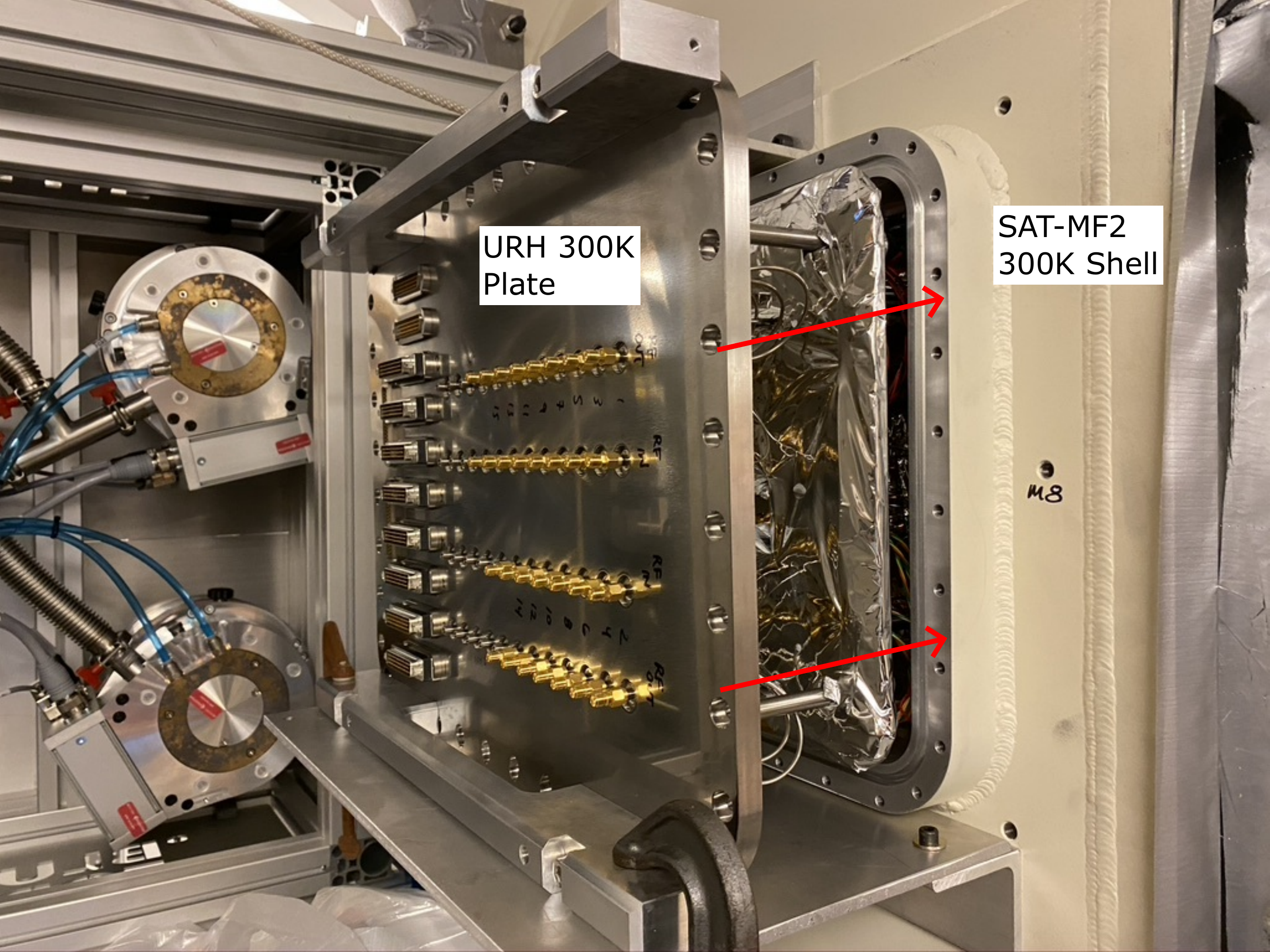}
    \caption{The URH during installation into the SAT-MF2 receiver.}
    \label{fig:tsatint}
\end{figure}

\begin{figure}
    \centering
    \includegraphics[width=0.9\textwidth]{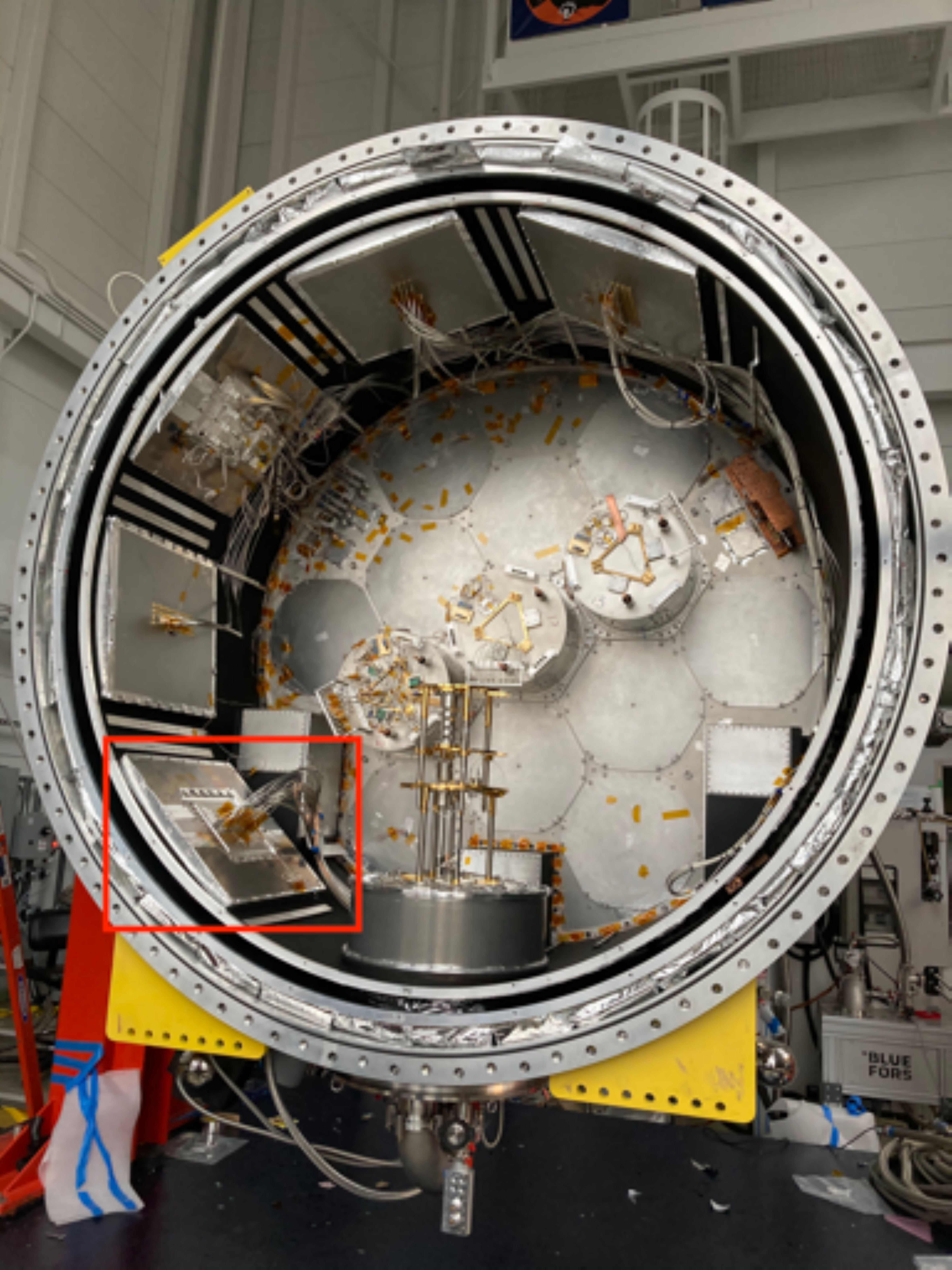}
    \caption{The Large Aperture Telescope Receiver. One of two integrated URHs is outlined in red. Photo courtesy of the Devlin Lab at UPenn.}
    \label{fig:latrint}
\end{figure}

\begin{figure}
    \centering
    \includegraphics[width=\textwidth]{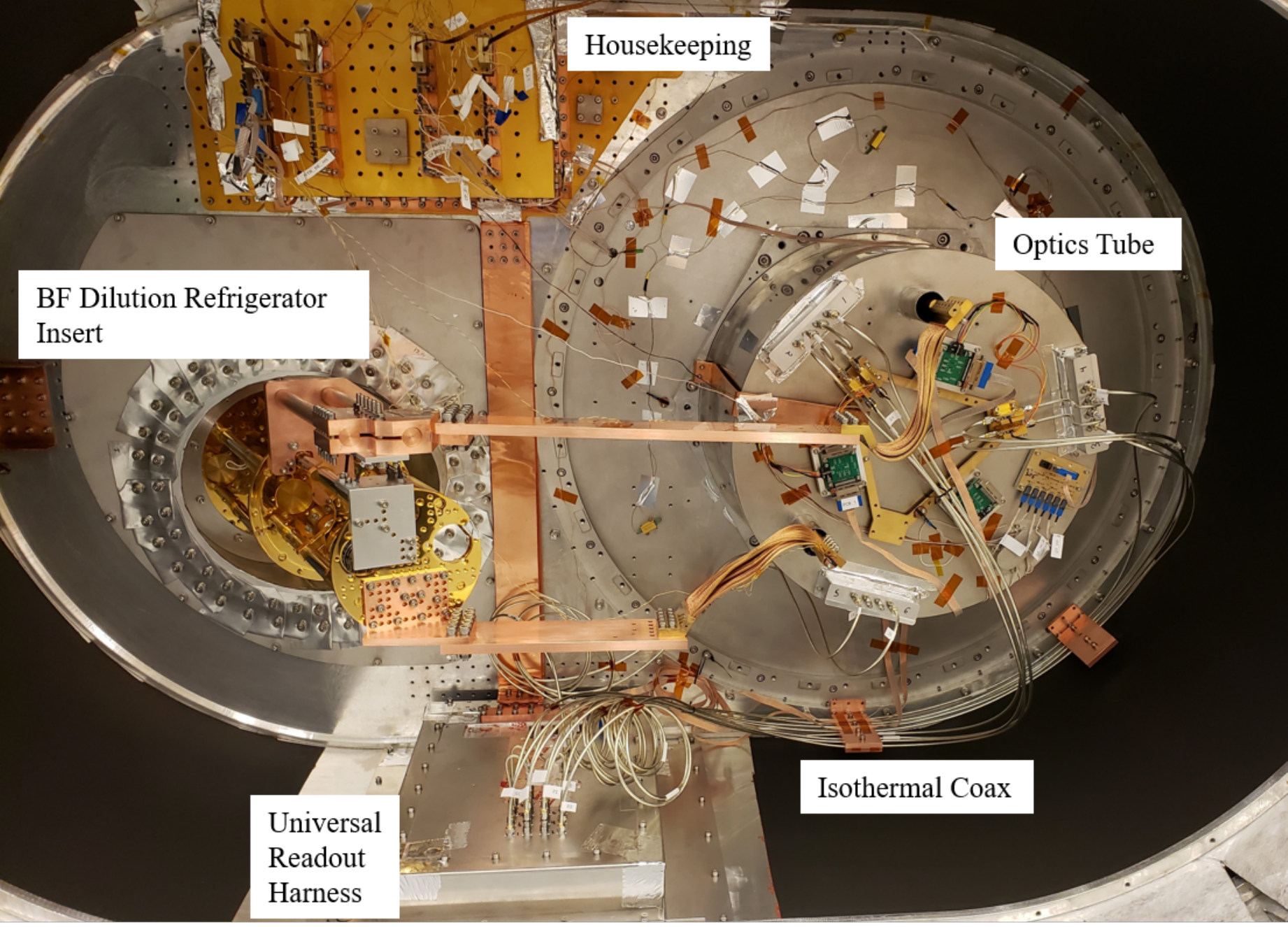}
    \caption{The LATR optics tube testbed with an integrated URH. Figure courtesy of Katie Harrington.}
    \label{fig:latrtint}
\end{figure}

Integration of harnesses into SO receivers is well underway. Thus far, two harnesses have been successfully integrated into the mid-frequency small aperture telescopes, SAT-MF1 at the University of California San Diego (Fig. \ref{fig:sat1int}) and SAT-MF2 at Princeton (Fig. \ref{fig:tsatint}), and a further two harnesses have been installed into the LATR at the University of Pennsylvania (Fig. \ref{fig:latrint}). A more sparsely populated (7 RF channel) harness has also been successfully integrated into the LATR ``tester'' (LATRt) at the University of Chicago, which has been used to validate the LATR optics tubes and will ultimately be converted into an additional SAT (Fig. \ref{fig:latrtint}).  Harness integration efforts for the ultra high-frequency SAT (SAT-UHF) at the University of California Berkeley remain ongoing. Both the LATR and SAT-1 are expected to deploy in the near future, and further details about post-integration readout performance for those receivers are presented in Refs.~\citenum{latr, latrt, annaspie, tanayspie} and in future publications \cite{satpaperinprep}.

\clearpage
\bibliography{bib} 
\bibliographystyle{spiebib} 

\end{document}